\documentclass[journal=jpccck,manuscript=article]{achemso}

\usepackage[version=3]{mhchem} 
\usepackage[utf8]{inputenc}
\usepackage{array}
\usepackage{wrapfig}
\usepackage{multirow}
\usepackage{tabularx}

\author{Raphael M. Tromer}
\affiliation[State University of Campinas]
{Applied Physics Department, State University of Campinas, Campinas, SP, 13083-970, Brazil}
\alsoaffiliation[State University of Campinas]
{Center for Computational Engineering and Sciences, State University of Campinas, Campinas, SP, 13083-970, Brazil}

\author{Levy Felix}
\affiliation[State University of Campinas]
{Applied Physics Department, State University of Campinas, Campinas, SP, 13083-970, Brazil}
\alsoaffiliation[State University of Campinas]
{Center for Computational Engineering and Sciences, State University of Campinas, Campinas, SP, 13083-970, Brazil}
\author{Cristiano F. Woellner}
\affiliation[Federal University of Parana]
{Physics Department, Federal University of Parana, UFPR, Curitiba, PR, 81531-980, Brazil}
\author{Douglas S. Galvao}
\email{galvao@ifi.unicamp.br}
\affiliation[State University of Campinas]
{Applied Physics Department, State University of Campinas, Campinas, SP, 13083-970, Brazil}

\title{A DFT Investigation of the Electronic and Optical Properties of Pentadiamond}

\usepackage{xcolor}
\begin{document}


\begin{abstract}
Recently, a new carbon 3D carbon allotrope named pentadiamond was proposed. Pentadiamond is composed of carbon atoms in mixed sp$^2$ and sp$^3$-like hybridization. In this work, we have carried out a detailed investigation of the electronic and optical properties of pentadiamond structure using first-principles (DFT) methods. Our results show that pentadiamond has an indirect bandgap semiconductor of $2.50$ eV with GGA-PBE and $3.31$ eV with HSE06. Its static dielectric constant is $4.70$ and the static refractive index is $2.16$. Pentadiamond presents low reflectivity, almost 40$\%$, for all-optical spectrum, making it a good structure to be used as a UV collector. Also, pentadiamond exhibits optical activity in the UV range where other carbon allotropes, such as diamond and 8-tetra(2,2) tubulane show no activity.

\end{abstract}

\section{Introduction}
                                                                                                                   
Carbon is capable of forming many 1D, 2D, or 3D allotropes as it easily form single, double and triple bonds \cite{benedek_2003,vanderbilt_1992,moras_2011,andrade_2015,robertson_2002,mackay_1991,Wang2015,Felipe}. In the nature, there are two natural allotropes forms, graphite and diamond, composed by sp$^3$ and sp$^2$ hybridized carbon atoms, respectively. Diamond has interesting physical properties such as high thermal conductivity \cite{Kidalov2009}, hardness \cite{diamond_hardness} and large optical absorption in the ultra-violet range \cite{diamond_optics}. Graphite is extremely soft \cite{gra_soft}, inert when in contact with a large  variety of materials \cite{gra_inert} and can be cleaved at very low pressure \cite{gra_cleave}. In 2004, a single graphite layer, the so-called graphene, was obtained using a mechanical exfoliation method \cite{Novoselov2004} and started a revolution in materials science \cite{Gwon2011}.

 However, pristine graphene is a semi-metal with a zero bandgap, which prevents its use in some architecture based on-off current regime. For this reason, much effort has been devoted to transforming graphene into a semiconductor but preserving some of its unique electronic properties. This has been tried using mechanical strain\cite{Wei2011}, external electrical field \cite{Castro2007}, and/or physical and chemical doping \cite{Dai2010}.
 
 The advent of graphene also renewed the interest for new 3D allotropes. Among the new proposed structures we can mention "protomene" \cite{PROTO_2018,PROTO_2019} and "novamene" \cite{NOVAMENE_2017,NOVAMENE_2018}, which are 3D allotropes composed by carbon in sp$^2$ and sp$^3$-like hybridization that presents a bandgap value of $3.0$ and $0.3$ eV, respectively.
 
 More recently, a new 3D carbon allotrope was theoretically proposed, "pentadiamond" (see Figure \ref{fig:structures}), an extended diamond form with sp$^2$ and sp$^3$ hybridizations \cite{PENTADIAMOND}. Besides the new topological features, in contrast to diamond which is an insulating material, pentadiamond is a semiconductor with an indirect bandgap value of $2.52$ eV \cite{PENTADIAMOND}. It was proposed \cite{PENTADIAMOND} that it was an auxetic material (negative Poisson's ratio), but it was a numerical mistake \cite{comment1,comment2}.
 
 In this work, we carried out a detailed investigation of the electronic and optical properties of the pentadiamond using Density Functional Theory (DFT) methods. For comparison purposes we also considered the diamond and the 8-tetra(2,2)-tubulane \cite{tubulane,tubulane2}, which is another 3D carbon allotrope containing sp$^2$ and sp$^3$ carbon-like hybridizations and a very close bandgap value (2.52 eV).

\section{Methodology}

The calculations in this work were performed using density functional theory (DFT) methods within the generalized gradient approximation (GGA) and the Perdew-Burke-Ernzerhof (PBE) functional for the exchange-correlation part \cite{Perdew1996}. We use a mesh cutoff energy value of $300$ Ry and a Brillouin zone sampling with $10\times 10\times 10$ set of $k$ point within Monkhorst-Pack scheme \cite{Monkhorst_1976}. We use the SIESTA software that it has all packages required for this purpose \cite{Soler2002}. 

When the total energy difference of successive iteration is smaller than $10^{-6}$, we assume the self-consistent field (SCF) convergence criteria were satisfied. The geometrical structural optimizations of the pentadiamond, diamond, and 8-tetra(2,2) structures were carried out using the conjugate gradient method. No constraints were used, we allow both atomic positions and lattice parameters to fully relax. Both atomic positions and lattice vectors are optimized simultaneously. We assume that the optimization convergence process is satisfied when forces on each atom are smaller than 0.005 eV/\AA.      

Once structures were optimized as discussed above, we performed the optical analysis in the linear regime considering an external electric field of magnitude $1.0$ \cite{Fadaie2016}. All optical calculations were also performed with the SIESTA software. We assume that external electrical field is polarized as an average of the three spatial directions, along x, y, and z-directions.

We define the complex dielectric function as:
\begin{equation}
\epsilon(\omega)=\epsilon_1(\omega)+i\epsilon_2(\omega) ,
\label{eq:dielectric_fun}
\end{equation}
where $\epsilon_1$ and $\epsilon_2$ are the real and imaginary parts of the dielectric function, respectively. 

The imaginary part of dielectric function $\epsilon_2$ can be extracted from Fermi's golden rule through interbands transitions \cite{Soler_2002}:
\begin{equation}
\epsilon_2(\omega)=\frac{4\pi^2}{\Omega\omega^2}\displaystyle\sum_{i\in \mathrm{VB},j\in \mathrm{CB}}\displaystyle\sum_{k}W_k|\rho_{ij}|^2\delta	(\epsilon_{kj}-\epsilon_{ki}-\omega).
\end{equation}
where $W_k$ is the individual k point weight, $\rho_{ij}$ is the dipole transition operator projected on the atomic orbitals basis with elements $i$ and $j$, $\Omega$ is the unit cell volume, $\omega$ is the photon frequency and VB and CB refer to the valence and conduction bands, respectively. 

From Kramers-Kronig relation we determine the real part of dielectric function:

\begin{equation}
\epsilon_1(\omega)=1+\frac{1}{\pi}P\displaystyle\int_{0}^{\infty}d\omega'\frac{\omega'\epsilon_2(\omega')}{\omega'^2-\omega^2},
\end{equation}
where $P$ denotes the principal value.

All optical quantities of interest such as the absorption coefficient $\alpha$, the reflectivity $R$, and the refractive index $\eta$,
can be evaluated directly from real and imaginary parts of dielectric function:

\begin{equation}
\alpha (\omega )=\sqrt{2}\omega\bigg[(\epsilon_1^2(\omega)+\epsilon_2^2(\omega))^{1/2}-\epsilon_1(\omega)\bigg ]^{1/2},
\end{equation}
\begin{equation}
\eta(\omega)= \frac{1}{\sqrt{2}} \bigg [(\epsilon_1^2(\omega)+\epsilon_2^2(\omega))^{1/2}+\epsilon_1(\omega)\bigg ]^{2}
\end{equation}

\begin{equation}
R(\omega)=\bigg [\frac{(\epsilon_1(\omega)+i\epsilon_2(\omega))^{1/2}-1}{(\epsilon_1(\omega)+i\epsilon_2(\omega))^{1/2}+1}\bigg ]^2 ,
\end{equation}

The accuracy of the optical calculation depends directly on the accuracy of the bandgap value. In the literature, is well-known that GGA-PBE in general underestimates the bandgap value for semiconductor materials \cite{Johnson_1998}. A comparison between GGA and other approximations, suggests that HSE06 functional produces bandgap values with higher precision\cite{Kishore2017}. As HSE06 is unavailable in the SIESTA software,  our calculation for bandgap value and optical transitions are underestimated. However, in SIESTA there is the possibility to carry out the optical calculations with corrections if the bandgap value was obtained with a more robust method, such as HSE06.  
The correction is introduced in SIESTA by a definition of the scissor operator, which produces a shift in the unoccupied states, given by,

\begin{equation}
\mathrm{scissor}=E_{gap}^{\mathrm{HSE06}}-E_{gap}^{\mathrm{PBE}}.
\label{scissor}
\end{equation}
In some works in the literature the corrected optical spectra obtained with the scissor operator produced results comparable to more sophisticated methods as as GW ones\cite{Nayebi2016,Ljungberg2017,Kolos2019}. 

Here, we used the Gaussian16 software package for obtaining the corrected bandgap value with the HSE06 functional. Then we used the scissor operator from equation \ref{scissor} for performing the optical calculations with the same accuracy presented by HSE06.

\section{Results}
\subsection{Structural Parameters}
\begin{figure}[ht]
\begin{center}
\includegraphics[width=0.8\linewidth]{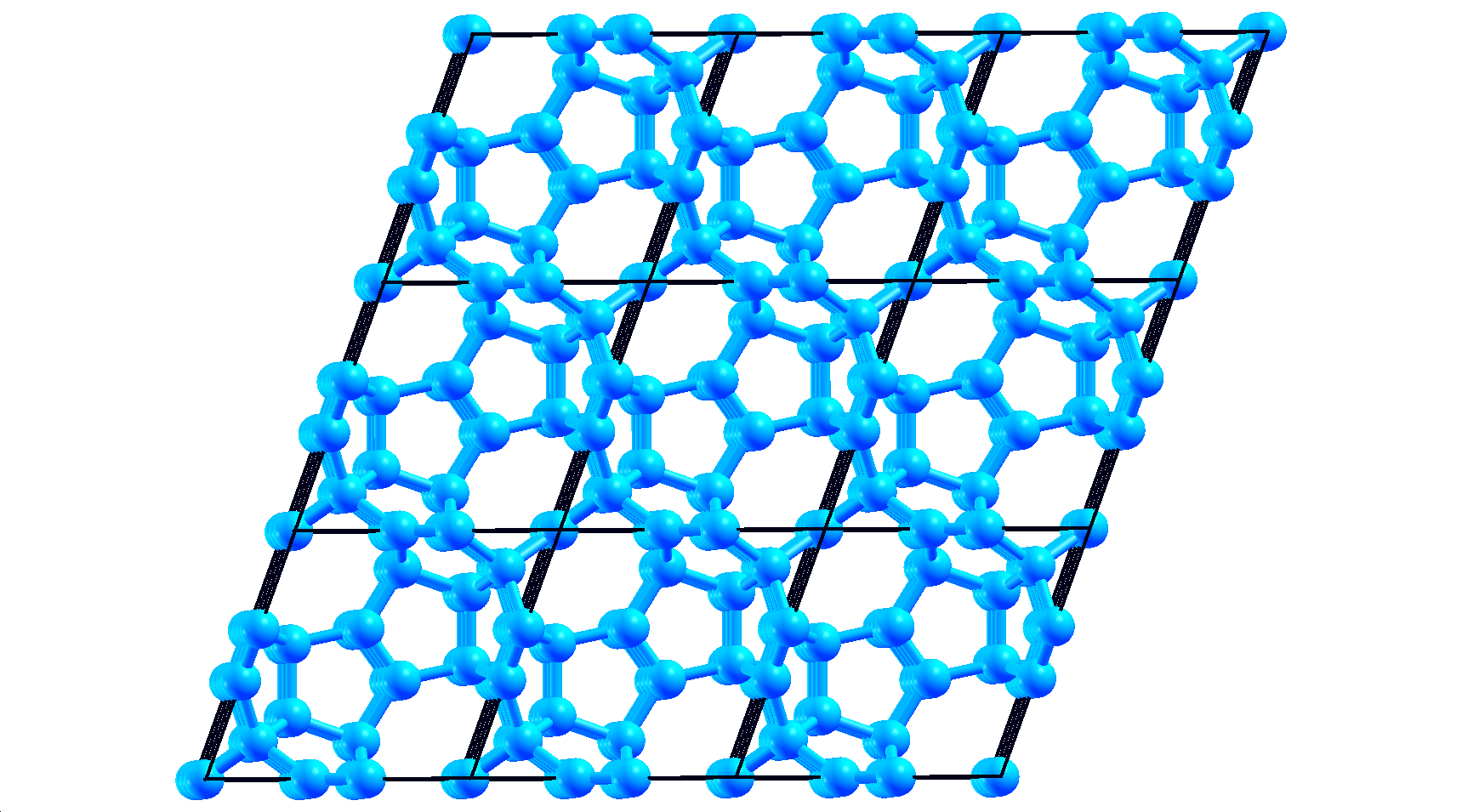}\\
\includegraphics[width=0.8\linewidth]{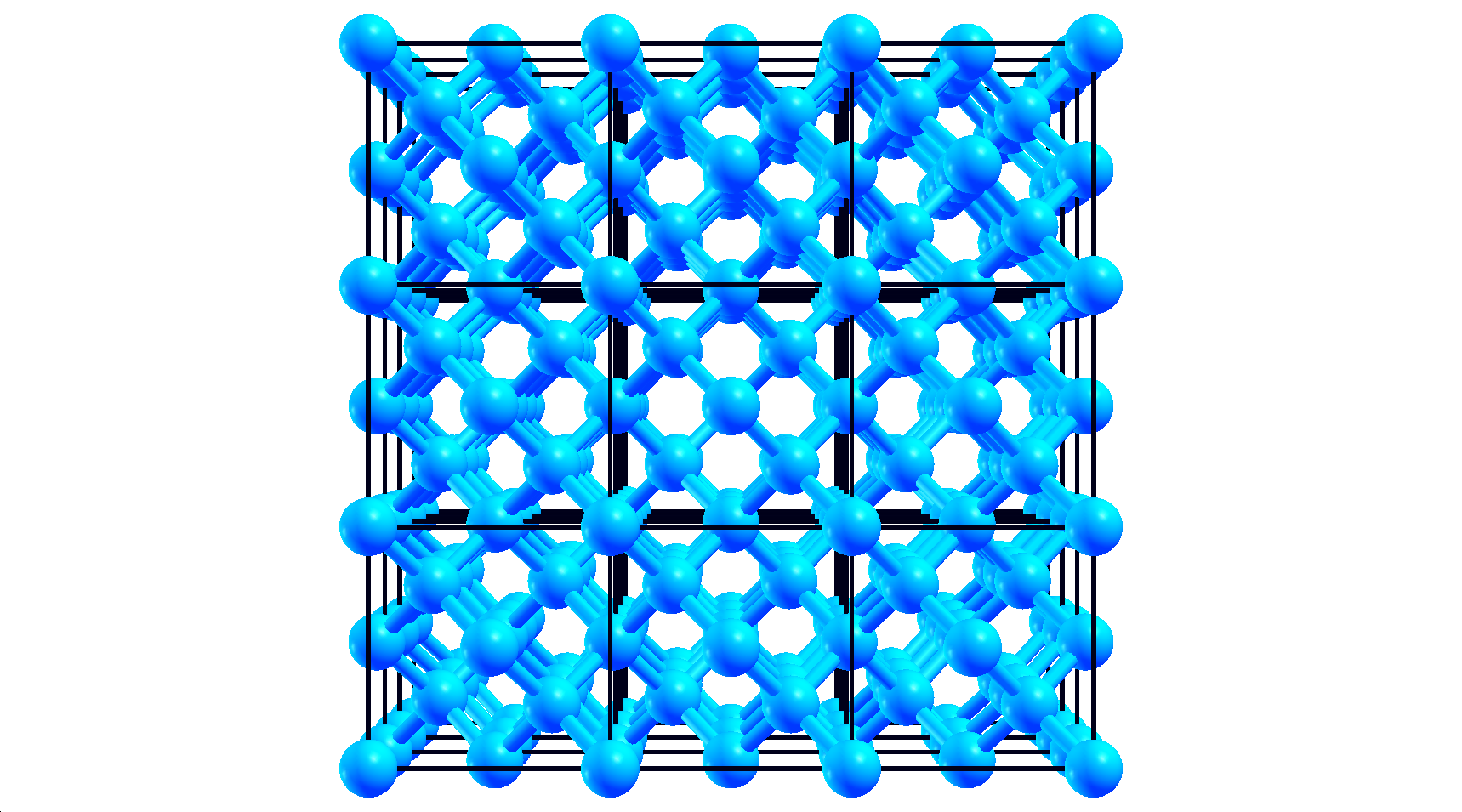}
\includegraphics[width=0.8\linewidth]{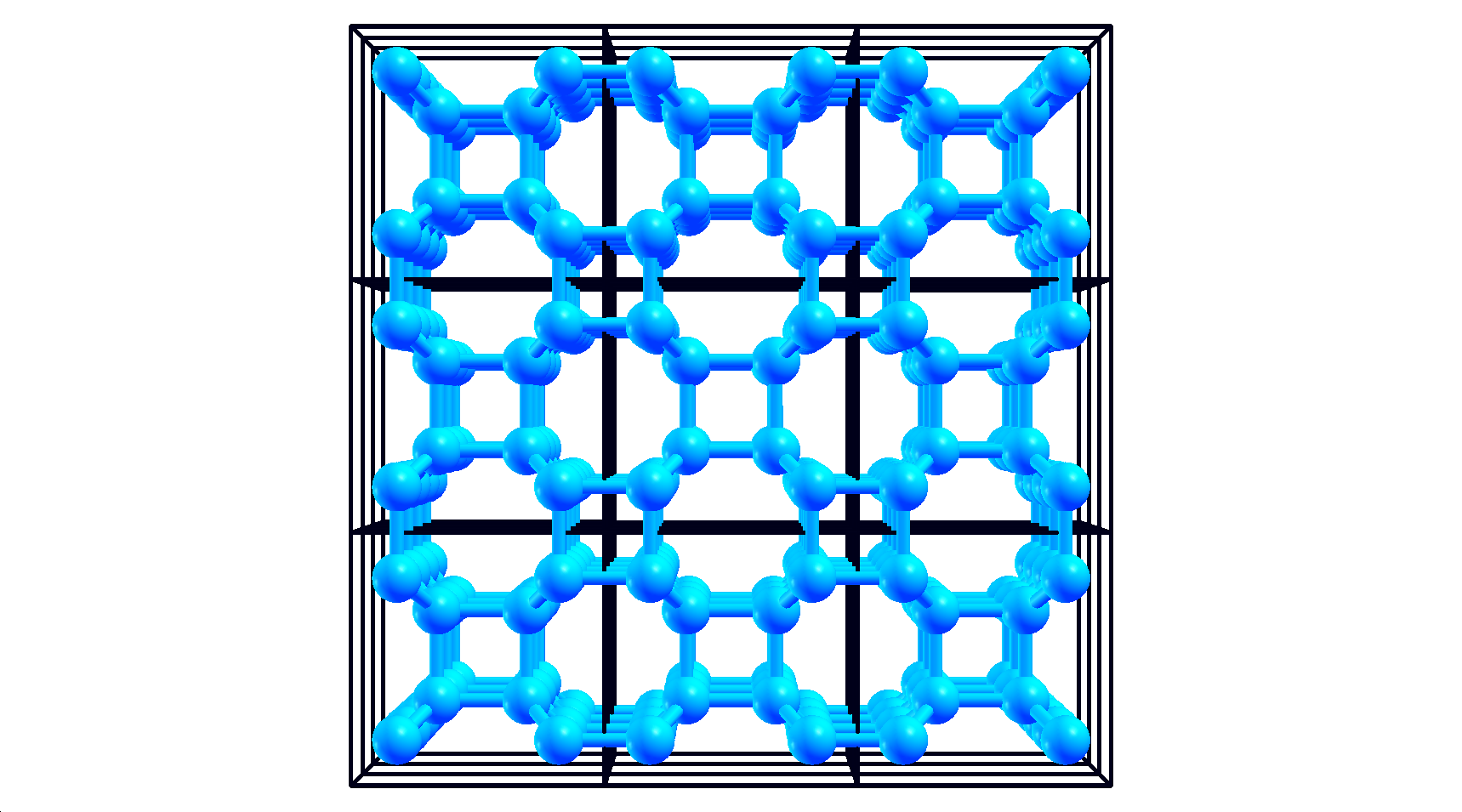}
\caption{\label{fig:structures} From top to bottom the optimized structures replicated $3\times 3\times 3$ along X, Y and Z of pentadiamond, diamond and 8-tetra(2,2), respectively. The black lines indicated the lattice vectors. The structural information is displayed in table \ref{tab:structures}.}
\end{center}
\end{figure}

In Figure \ref{fig:structures} we present the optimized structures replicated $3\times 3\times 3$ along X, Y and Z of the structures investigated in this work: pentadiamond, diamond, and 8-tetra(2,2). The black lines indicate the lattice vectors.

As mentioned above, pentadiamond and 8-tetra(2,2) consist of carbon in sp$^2$ and sp$^3$-like hybridization, while diamond contains sp$^3$ carbons. The average C-C bond length for each case were $1.51$, $1.56$ and $1.54$~\AA~for pentadiamond, diamond and 8-tetra(2,2), respectively. These results are expected  because sp$^3$ bond-lengths are more larger than sp$^2$ ones, therefore as pentadiamond and 8-tetra(2,2)  contains sp$^2$-like carbons their average C-C  bond-lengths is to be smaller, as observed. The optimized structural parameters are presented in Table \ref{tab:structures}. It should be stressed that pentadiamond and diamond are fully isotropic while 8-tetra(2,2) only along the X and Y axes. Also, the pentadiamond primitive supercell is FCC while diamond and 8-tetra(2,2) are tetragonal.  

\begin{table}[ht]
\begin{center}
\begin{tabular}{|c|c|c|c|}
\hline
Parameters&Pentadiamond&Diamond&8-tetra(2,2)\\
\hline
atoms per unit cell&22&8&8\\
\hline
a (\AA)&6.55&3.56&4.39\\
\hline
b (\AA)&6.55&3.56&4.39\\
\hline
c (\AA)&6.55&3.56&2.52\\
\hline
$\alpha~(^o)$&60.0&90.0&90.0\\
\hline
$\beta~(^o)$&60.0&90.0&90.0\\
\hline
$\gamma~(^o)$&60.0&90.0&90.0\\
\hline 
Volume (\AA$^3$)&199.13&44.97&48.54\\
\hline
Density (g/cm$^3$)&2.20&3.54&3.30\\
\hline
Space Group Name (number) &Fm${\bar{3}}$m (225)& Fd${\bar{3}}$m (227)& I4/mmm (139)\\
\hline
\end{tabular}    
\end{center}
\caption{\label{tab:structures} Structural information from GGA-PBE calculations for pentadiamond, diamond and 8-tub(2,2).}
\end{table}

\subsection{Electronic Analysis}
\begin{figure}[ht]
\begin{center}
\includegraphics[width=0.8\linewidth]{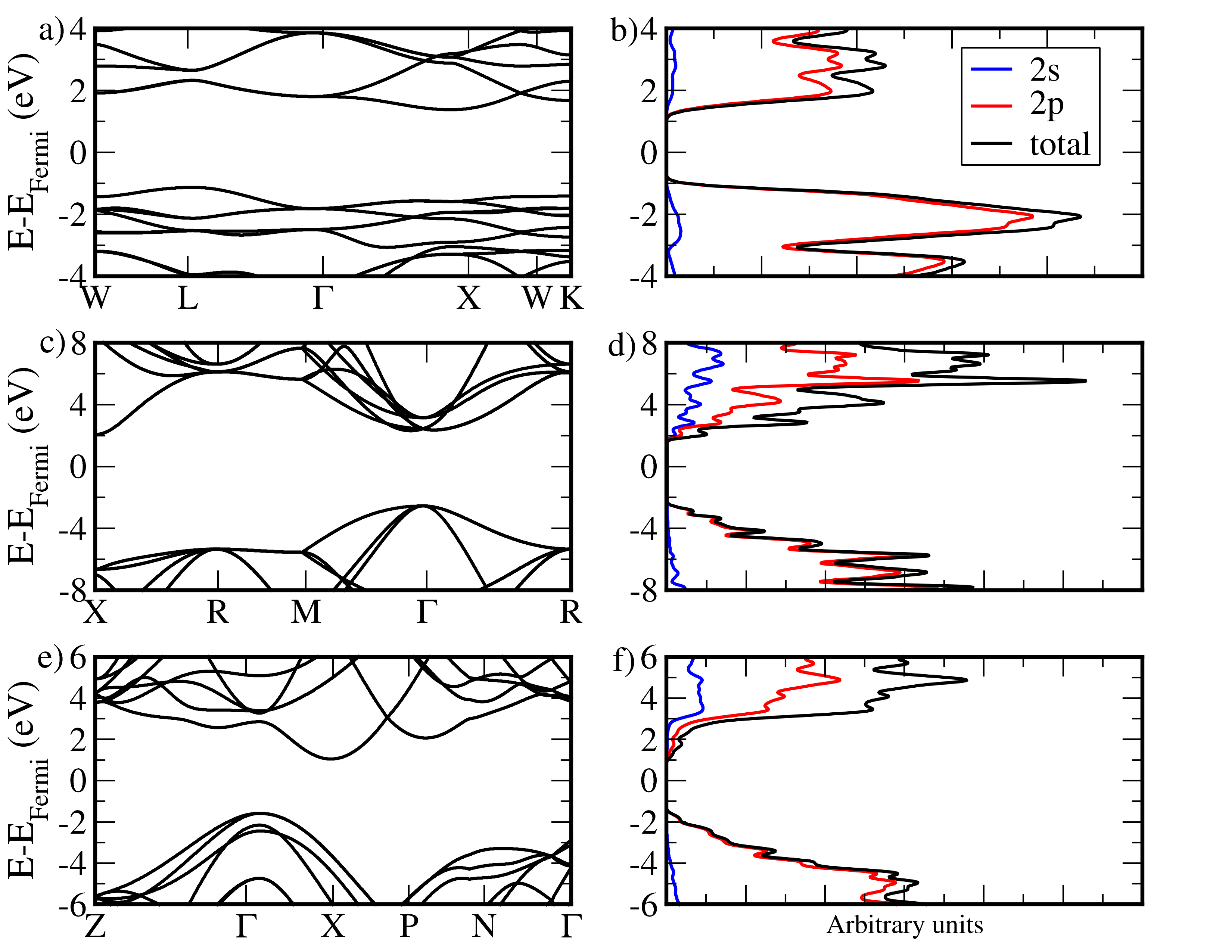}\\
\caption{\label{fig:bands} Electronic band structure and their corresponding projected density of states (PDOS) for: a/b) pentadiamond, c/d)diamond, and e/f) 8-tetra(2,2).}
\end{center}
\end{figure}

In Figure \ref{fig:bands} we present the pentadiamond, diamond, and 8-tetra(2,2) band structure and corresponding projected density 
of states (PDOS) for valence atomic orbitals $2s$ and $2p$. From the band structure, we observe that pentadiamond and 8-tetra(2,2) structures are semiconductor materials with bandgap 
values of $2.50$ and $2.65$ eV,respectively, while the diamond has a bandgap values of $4.20$ eV. We observe an indirect bandgap in all cases, the bandgap values are given by energy difference from frontier orbitals levels highest occupied crystal orbital (HOCO) and lowest unoccupied crystal orbital (LUCO). The frontier orbital HOCO are at the symmetry point $L$, $\Gamma$ and $\Gamma$, for pentadiamond, diamond and 8-tetra(2,2), respectively, while the frontier orbital LUCO is 
at $X$ point, for all structures. We observe from PDOS that frontier orbtials HOCO and LUCO are predominantly composed of atomic orbitals $2p$ of carbon atoms. 

We mentioned early in the methodology section the bandgap values obtained using GGA-PBE is underestimated. For the optimized geometries shown in Figure \ref{fig:structures}, we performed one self-consistent field calculation with gaussian16 software using the HSE06 functional and basis set cc-pVTZ in order to obtain more realistic bandgap values. Using the experimental bandgap value of diamond as the reference ($5.5$ \cite{Wort2008}) HSE06-cc-PVTZ produced the best result ($5.4$ eV), very close to the experimental value. For pentadiamond and 8-tetra(2,2) we obtained $3.31$ and $3.76$ eV, respectively (in comparison to $2.50$ and $2.65$ eV, respectively, obtained with GGA-PBE).

\subsection{Optical Properties}

To perform the optical calculations we used the
scissor correction to produce a more precise description of the optical
transitions, as mentioned in the methodology section. In the last section, we obtained the bandgap values through gaussian16 with the HSE06 functional. Since the obtained bandgap value for the diamond structure is very close to the experimental one and all structures are composed only by carbon atoms, we can expect that the values for pentadiamond and 8-tetra(2,2) were also obtained with good accuracy. There are no calculations in the literature for these structures using a hybrid functional such as HSE06. With the scissor operator correction our results should be equivalent to those performed with the HSE06 functional. 

\begin{figure}[ht]
\begin{center}
\includegraphics[width=0.8\linewidth]{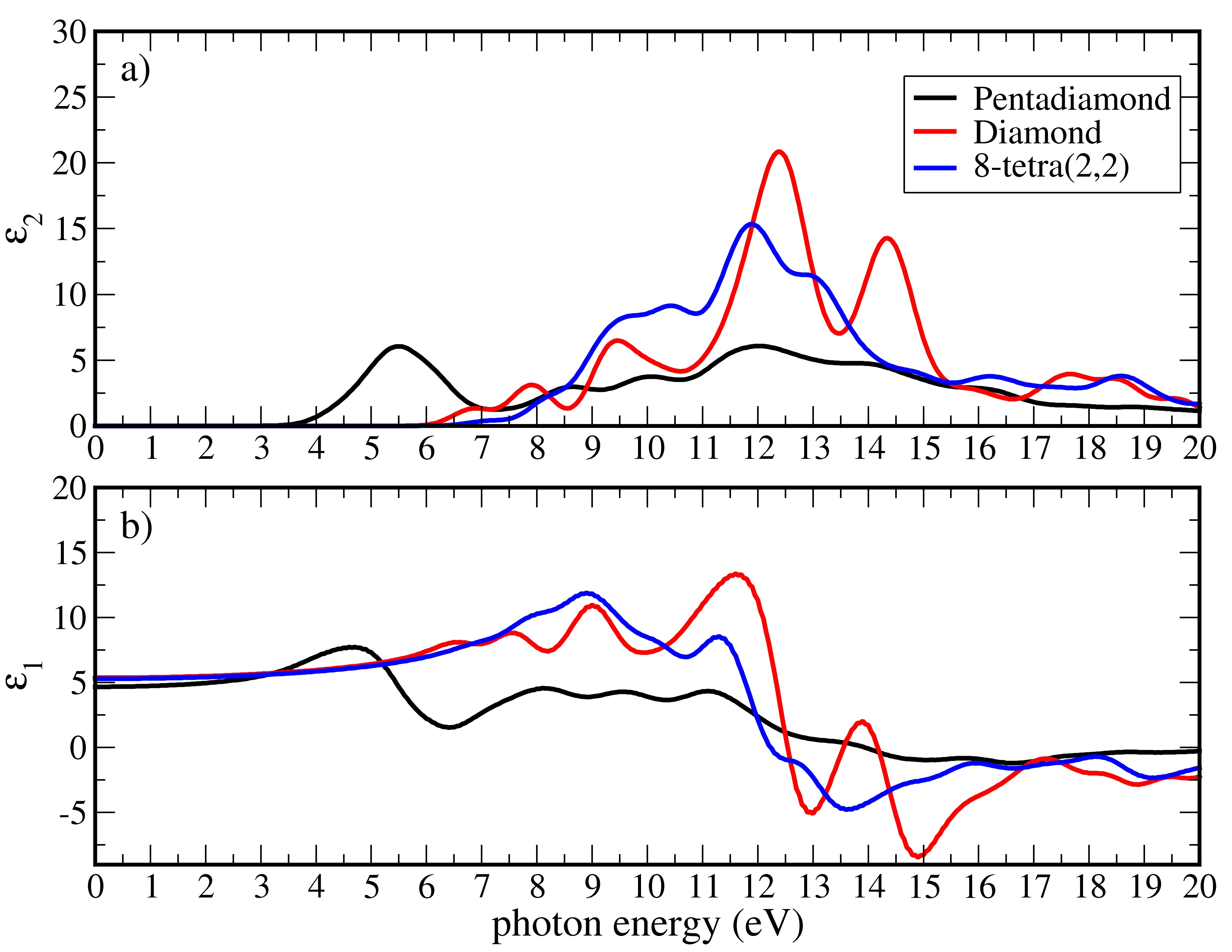}\\
\caption{\label{fig:epsilon} Imaginary a) and real b) part of dielectric constant function as a function of photon energy. }
\end{center}
\end{figure}

In Figure \ref{fig:epsilon}, we present the imaginary a) and real b) part of dielectric constant function as a function of photon energy value. The imaginary and real part of dielectric constant show the absorptive and dispersion behaviour, respectively. The external electrical field is polarized as an average along X, Y and Z directions.
Each peak in Figure \ref{fig:epsilon}-a) corresponds to an optical transition. We notice that imaginary part of dielectric constant function starts for nonzero values close to the 3.5 eV for the pentadiamond. This is consistent with the corrected bandgap values. For this reason we do not observe optical activity in
infrared region. The materials are transparent in this region and does not occur attenuation of the incident light. We observe the the optical transition only occur for photon energy corresponding to the ultra violet region. Also, pentadiamond and diamond start to absorb when photon energy near their bandgap values. These transitions occur involving the symmetry points $X$ and $K$ on the pentadiamond and $\Gamma$ on the diamond.  In contrast, for 8-tetra(2,2) we observe the direct transition at $\Gamma$ and $N$, which are associated with energies significantly larger than bandgap value. We observe that for photon energies larger than $7$ until $20$ eV, diamond and 8-tetra(2,2) tend to absorb more than pentadiamond.

In Figure \ref{fig:epsilon}-b) we present the real part of the dielectric constant function as a function of the photon energy. We did not observe light attenuation for photon energies lower than $3.5$ eV. As mentioned early, there is no optical activity in the infrared range. The dielectric constant value can be extracted from Figure \ref{fig:epsilon}-b) at long wavelength regime (photon energy tend to zero). The obtained values for dielectric constant: $4.70$, $5.40$ and $5.43$ for pentadiamond, diamond ($5.6$ is the experimental value, from \cite{Bangert2006}) and 8-tetra(2,2), respectively. The dielectric constant value is important for the electronic transport properties, as it affects the exciton (electron-hole pairs) separation.

\begin{figure}[ht]
\begin{center}
\includegraphics[width=0.8\linewidth]{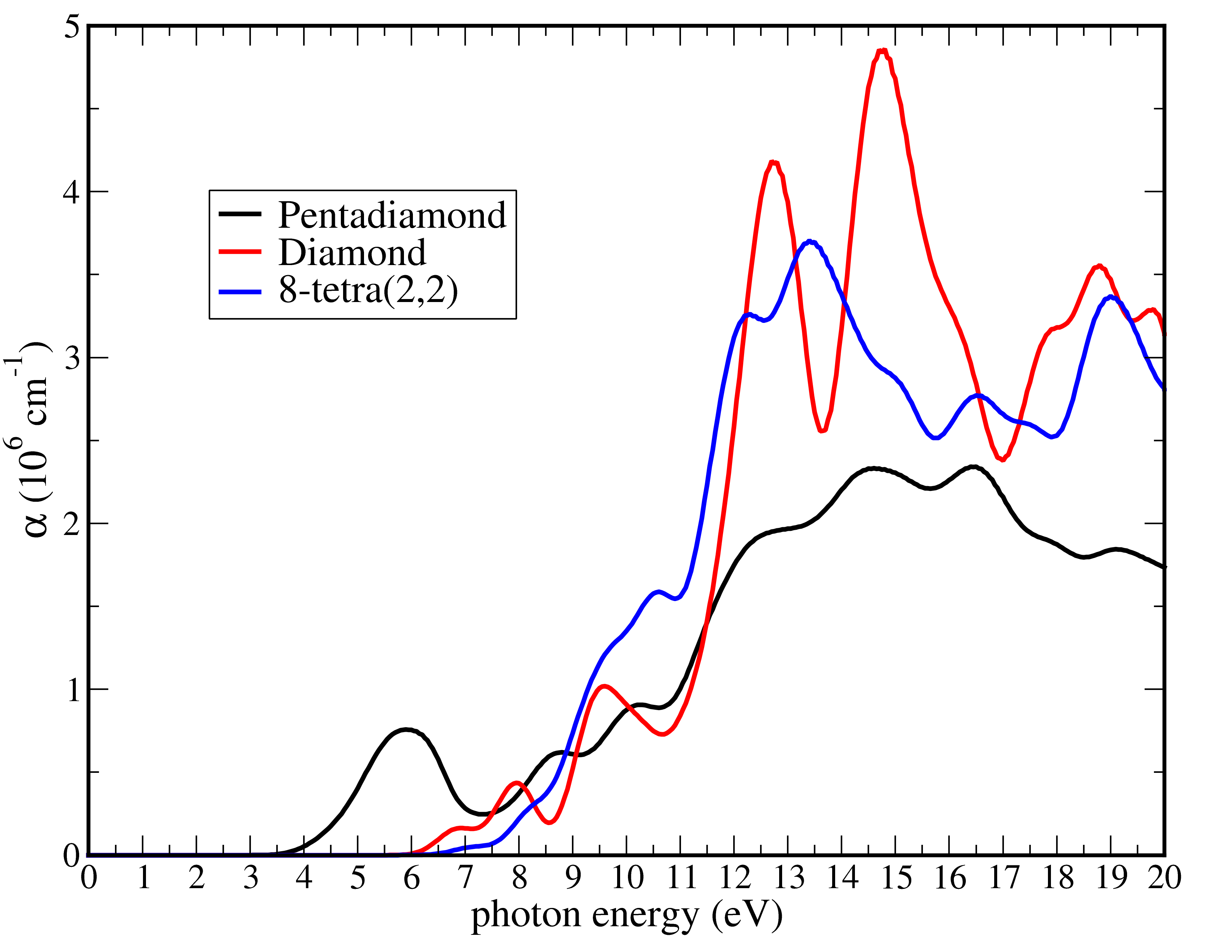}\\
\caption{\label{fig:abs} Absorption coefficient $\alpha$ as a function of photon energy. }
\end{center}
\end{figure}

In Figure \ref{fig:abs} we show the absorption coefficient as a function of photon energy. All peaks are located in the ultra-violet region because the structures investigated here present large bandgap values. For photon energies values lower than $7$ eV, the absorption intensity is more strong for pentadiamond. From $7$ eV we observe for all cases that intensities increase until the two almost equivalent (in intensity) peaks located at $14.0$ and $16.8$ eV for pentadiamond, $14.8$ eV for the diamond, and $13.4$ eV for 8-tetra(2,2). The maximum absorption intensities are $2.3\times 10^6$cm$^{-1}$, $4.8\times 10^6$cm$^{-1}$ and $3.5\times 10^6$cm$^{-1}$, for pentadiamond, diamond and 8-tetra(2,2), respectively. For carbon carbon, in general, intense absorption in UV region is associated with sp$^3$-like carbons. In this sense, the sp$^3$ content decrease from diamond, 2-tetra(2,2), to pentadiamond, which follows the intensity ordering of Figure \ref{fig:abs}.

\begin{figure}[ht]
\begin{center}
\includegraphics[width=0.8\linewidth]{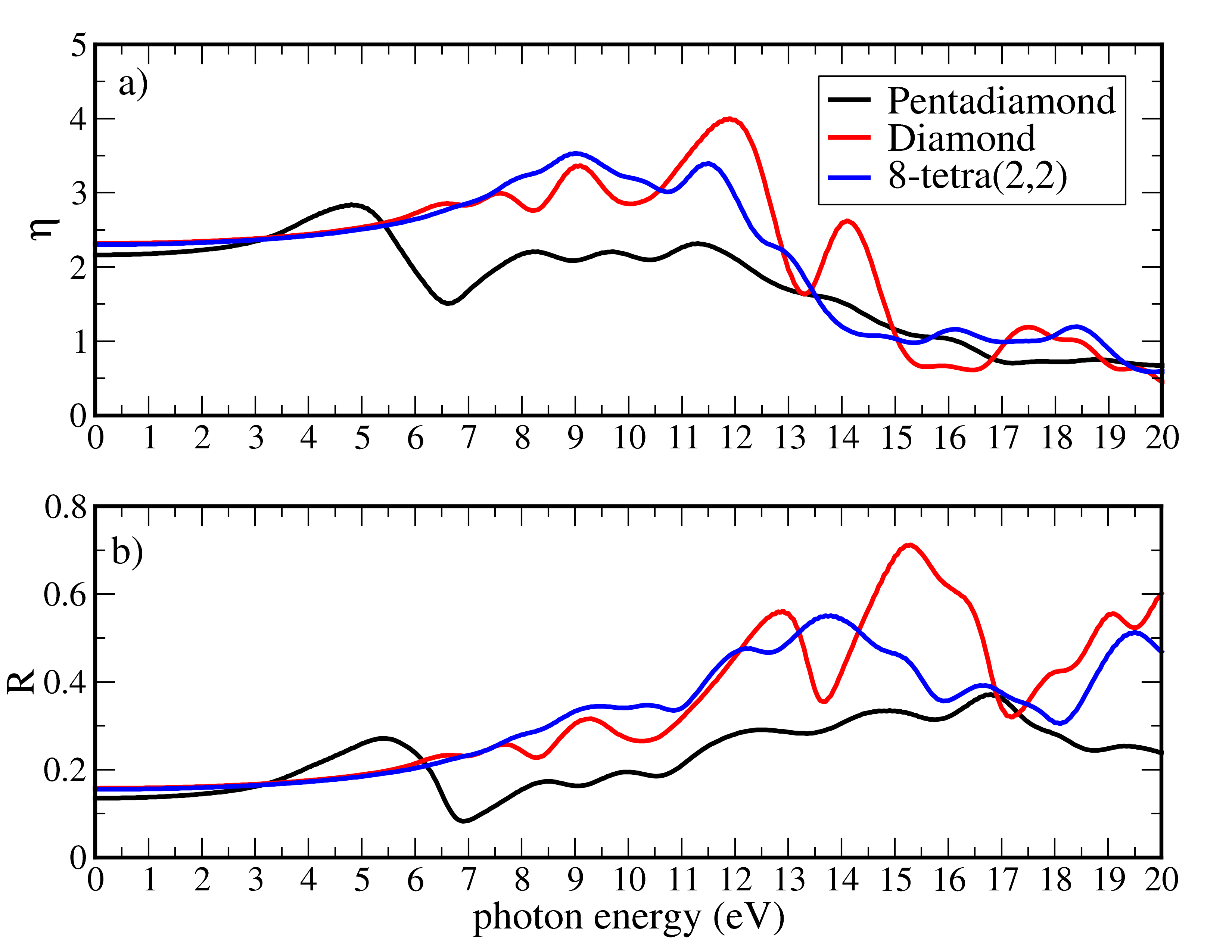}\\
\caption{\label{fig:ref} Refractive index a) and reflectivity b) as a function of photon energy.}
\end{center}
\end{figure}

In Figure \ref{fig:ref} we show the refractive $\eta$ a) and reflectivity R b) indices as a function of photon energy. As mentioned before, we do not observe light attenuation for photon energies until $7$ eV for diamond and 8-tetra(2,2) cases. The maximum peak for each case in Figure \ref{fig:ref}-a) are $5$, $11.9$ and $9.0$ eV for pentadiamond, diamond and 8-tetra(2,2), respectively. In the range from $7$ to $12.8$ eV the refractive index for diamond and 8-tetra(2,2) are higher than the one for pentadiamond. For higher photon energies values all systems exhibit the same trends. The static refractive index value of the material can be extracted from Figure \ref{fig:ref}-a) in the long wave regime (photon energy tend to zero).   

The obtained values are $2.16$, $2.33$ and $2.30$ for pentadiamond, diamond ($2.4$ for the experimental value, from \cite{Kim2013}) and 8-tetra(2,2), respectively. 

In Figure \ref{fig:ref}-b) we can observe the that reflectivity of pentadiamond has the maximum peak at $17$ eV, where it reflects almost $40\%$ of UV incident light, while diamond reflects $70\%$ at $15.2$ eV and 8-tetra(2,2) reflects almost $60\%$ at $13.7$ eV. Therefore, pentadiamond can be used as UV collector for photon energies until $15$ eV, because of its low reflectivity and refractive index greater than 1. In contrast, diamond and 8-tetra(2,2) can be used as UV collector only until $11$ eV, from this value their reflectivity increase and the refractive indices decrease.

\section{Conclusion}

In summary, we have carried out a detailed investigation of the electronic and optical properties of pentadiamond structure using first-principles (DFT) methods. Pentadiamond is a carbon allotrope structure composed of mixed of sp$^2$ and sp$^3$-like hybridization. For comparison purposes, we have also considered two other structures: diamond and 8-tetra(2,2)tubulane. Our results show that pentadiamond has an average equilibrium bond-length distance of $1.51$ \AA ($1.54$ and $1.56$ for diamond and 8-tetra(2,2), respectively). Pentadiamond is an indirect bandgap semiconductor ($2.50$ eV with GGA-PBE and $3.31$ eV with HSE06). The static dielectric constant for pentadiamond is $4.70$. The maximum absorption intensities are $2.3\times 10^6$cm$^{-1}$, $4.8\times 10^6$cm$^{-1}$ and $3.5\times 10^6$cm$^{-1}$, for pentadiamond, diamond and 8-tetra(2,2), respectively. The static refractive index for pentadiamond is $2.16$ Pentadiamond presents low reflectivity, almost 40$\%$, for all-optical spectrum. From our analysis, we conclude that pentadiamond can be used as a UV collector because of its low reflectivity. Also, pentadiamond exhibits optical activity in the UV range where other structures as diamond and 8-tetra(2,2) show no activity.

\section{Acknowledgements}

This work was financed in part by the Coordenacão de Aperfeiçoamento de Pessoal de Nível Superior - Brasil (CAPES) - Finance Code 001, CNPq, and FAPESP. The authors thank the Center for Computational Engineering \& Sciences (CCES) at Unicamp for financial support through the FAPESP/CEPID Grant 2013/08293-7. RMT also acknowledges support from the High Performance Computing Center at UFRN (NPAD/UFRN).

\pagebreak


\pagebreak
\bibliography{achemso-demo}

\end{document}